\documentclass[twocolumn,showpacs,amssymb,aps,prd,floatfix]{revtex4-1}
         
\usepackage{epsfig}
\usepackage{multirow}

\newcommand\tr{\,{\rm tr}\,}

\newcommand{\bra}{\langle}
\newcommand{\ket}{\rangle}

\newcommand{\id}{1\!\!1}

\newcommand{\half}{\frac{1}{2}}
\newcommand{\thalf}{\frac{3}{2}}

\newcommand{\vecp}{{\mathbf p}}

\newcommand{\pv}{{\mathbf p}}

\newcommand{\be}{\begin{equation}}
\newcommand{\ee}{\end{equation}}
\newcommand{\bea}{\begin{eqnarray}}
\newcommand{\eea}{\end{eqnarray}}
\newcommand{\bean}{\begin{eqnarray*}}
\newcommand{\eean}{\end{eqnarray*}}

\begin{document}

\title{Hyperons in thermal QCD: A lattice view}

\author{Gert Aarts$^a$}
\author{Chris Allton$^a$}
\author{Davide De Boni$^a$}
\author{Benjamin J\"ager$^b$}

\affiliation{$^a$Department of Physics, College of Science, Swansea University, Swansea SA2 8PP, United Kingdom}
\affiliation{$^b$CP$^3$-Origins \& Danish IAS, Department of Mathematics and Computer Science, University of Southern Denmark, 5230 Odense M, Denmark}

\date{Dated: December 18, 2018}

\begin{abstract}
The hadron resonance gas (HRG) is a widely used description of matter under extreme conditions, e.g.\ in the context of heavy-ion phenomenology. Commonly used implementations of the HRG employ vacuum hadron masses throughout the hadronic phase and hence do not include possible in-medium effects. Here we investigate this issue, using nonperturbative lattice simulations employing the FASTSUM anisotropic $N_f=2+1$ ensembles. We study the fate of octet and decuplet baryons as the temperature increases, focussing in particular on the positive- and negative-parity groundstates. While the positive-parity groundstate masses are indeed seen to be temperature independent, within the error, a strong temperature dependence is observed in the negative-parity channels. We give a simple parametrisation of this and formulate an in-medium HRG, which is particularly effective for hyperons. Parity doubling is seen to emerge in the deconfined phase at the level of correlators, with a noticeable effect of the heavier $s$ quark.
Channel dependence of this transition is analysed.
\end{abstract}


\maketitle


\section{Introduction}
\label{sec:intro}

How the light hadrons behave under the extreme conditions of nonzero temperature and/or density is a question of fundamental importance, linked to confinement and chiral symmetry. Moreover, quantitative insight is highly relevant for ongoing heavy-ion collision experiments, exploring the QCD phase diagram, as heavy-ion phenomenology relies on understanding the behaviour of hadrons under the conditions created in those collisions. 

A widely used description of the hadronic phase at nonzero temperature and vanishing or low baryon density is given by the hadron resonance gas (HRG) \cite{Dashen:1969ep,Venugopalan:1992hy,Karsch:2003vd}, in which all states, including resonances, contribute to thermodynamic quantities, such as pressure, entropy and susceptibilities. In its simplest form, the hadrons are treated as noninteracting particles, characterised by their vacuum mass and quantum numbers. This setup gives a reasonable description of both heavy-ion phenomenology \cite{Andronic:2005yp,Andronic:2017pug} and lattice QCD data \cite{Borsanyi:2010bp,Bazavov:2014pvz}. A closer look, however, reveals intriguing discrepancies \cite{Bazavov:2014xya,Alba:2017mqu}.
This is not unexpected and there are a number of HRG modifications which aim to go beyond the ideal HRG, e.g.\ by including attractive and/or repulsive interactions \cite{Vovchenko:2016rkn,Huovinen:2017ogf}, hard-core interactions \cite{Satarov:2016peb} or additional states, predicted by e.g.\ the quark model but not yet observed in nature \cite{Majumder:2010ik,Bazavov:2014xya,Alba:2017mqu}.

Most commonly used implementations of the  HRG, however, use a formulation in which the resonances have the same mass throughout the hadronic part of the QCD phase diagram, i.e.\ equal to their value in vacuum. This is somewhat surprising, since the in-medium modification of masses, or of more general spectral features, is a longstanding topic of interest in the theory and phenomenology of strongly interacting matter and often used as one of the indications of the presence of a medium \cite{Brown:1995qt,Rapp:1999ej}. 
For light hadrons, chiral symmetry is expected to play an important role and in-medium effects are often captured by a medium-dependent chiral condensate. While this is widely accepted, it is typically not implemented in the HRG or its extensions.
An exception is Ref.\ \cite{BraunMunzinger:2003zz}, in which the interplay between the freeze-out temperature and chiral-condensate dependent in-medium masses has been discussed.

There is therefore a need to unambiguously establish if and how the masses of the light hadrons in the hadronic phase depend on the temperature, at zero and low baryon density. This is a nonperturbative question in QCD, which can be addressed  using either a first-principle lattice QCD computation or via effective models, suitably benchmarked against lattice QCD results.
While mesons at finite temperature have indeed been fairly well studied on the lattice in various contexts (see e.g.\ Ref.~\cite{Aarts:2017rrl} for a list of references), for baryons this is not the case. In fact, there are only a few lattice studies of baryonic thermal screening \cite{DeTar:1987ar,Pushkina:2004wa} and temporal \cite{Datta:2012fz,Aarts:2015mma,Aarts:2017rrl} masses.  
Here we extend our work on baryons \cite{Aarts:2015mma,Aarts:2017rrl}, using lattice QCD simulations on the FASTSUM anisotropic $N_f=2+1$ ensembles \cite{Aarts:2014cda,Aarts:2014nba},  to include hyperons, i.e.\ all octet and decuplet baryons.
We are particularly interested in the role of chiral symmetry and the emergence of parity doubling: in the presence of unbroken chiral symmetry, positive- and negative-parity baryonic channels are degenerate. When chiral symmetry is (spontaneously) broken, this is no longer the case and indeed, in vacuum the positive-parity groundstate is typically lighter than the negative-parity one. Since chiral symmetry is restored around the deconfinement transition,  one expects a degeneracy to emerge, based purely on symmetry considerations. Here we investigate how this degeneracy emerges, which is a nonperturbative dynamical question, not answerable using symmetry considerations alone, and how it affects the HRG description.

The paper is organised as follows. In the following section we summarise the details of the lattice QCD simulations \cite{Aarts:2014cda,Aarts:2014nba}. Sec.\ \ref{sec:had} contains the results on temperature effects in the hadronic phase at vanishing baryon density, including a parametrisation of the temperature-dependent masses for the negative-parity groundstates. These parametrisations are subsequently used in Sec.\ \ref{sec:HRG} to define an in-medium hadron resonance gas, and we give an application to baryonic fluctuations and contributions to the pressure. Sec.\ \ref{sec:decon} contains the results in the deconfined phase, including an analysis of the strangeness dependence of the transition from the low- to the high-temperature phase. A summary is given in Sec.\ \ref{sec:conc}.
The $N, \Delta$ and $\Omega$ channels were already discussed in Ref.~\cite{Aarts:2017rrl}; preliminary results in the $\Lambda, \Sigma, \Sigma^*, \Xi$ and $\Xi^*$ channels have been presented in Ref.~\cite{Aarts:2017gmj}.


\section{Lattice QCD details}
\label{sec:lat}

In order to investigate QCD at nonzero temperature nonperturbatively, the FASTSUM collaboration uses lattice QCD simulations on highly anistropic lattices, $a_\tau/a_s \ll 1$. 
We employ a fixed-scale approach in which the temperature is varied by changing the number of time slices $N_\tau$, via the standard relation, $T=1/(a_\tau N_\tau)$.
The choice of fixed-scale anisotropic lattices is specifically motivated for spectral studies of QCD at nonzero temperature, as it increases the number of time slices available and does not require a zero-temperature tuning of parameters for each temperature value. 
Our lattice discretisation follows the Hadron Spectrum Collaboration  \cite{Edwards:2008ja}
 and uses a Symanzik-improved anisotropic gauge action with tree-level mean-field coefficients and a mean-field--improved Wilson-clover fermion action with stout-smeared links \cite{Morningstar:2003gk}. Full details can be found in Refs.~\cite{Aarts:2014cda,Aarts:2014nba}.

\begin{table}[t]
\centering
\begin{tabular}{c||cccc | cccc}
\hline
$N_\tau$ & 128 & 40 & 36 & 32 & 28 & 24 & 20 & 16 \\
\hline
$T$ [MeV] 	& 44 & 141  & 156   & 176 & 201  & 235  & 281 & 352 \\
$T/T_c$  		& 0.24 & 0.76  & 0.84   & 0.95 & 1.09  & 1.27  & 1.52 & 1.90 \\
$N_{\rm cfg}$ 	& 139  & 501 & 501 & 1000  & 1001 & 1001 & 1000 & 1001\\
$N_{\rm src}$ 	& 16  & 4 & 4 & 2 & 2 & 2 & 2 & 2 \\
\hline
\end{tabular}
\caption{Ensembles used in this work. The lattice size is $24^3 \times N_\tau$ , with the temperature $T = 1/(a_\tau N_\tau)$.
The available statistics for each ensemble is $N_{\rm cfg} \times N_{\rm src}$. The sources were chosen randomly in the four-dimensional lattice. The spatial lattice spacing $a_s = 0.1227(8)$ fm, the inverse temporal lattice spacing $a^{-1}_\tau = 5.63(4)$ GeV, and the renormalised anisotropy $\xi=a_s/a_\tau =3.5$.
The estimate $T_c=185(4)$ MeV is determined via the inflection point of the renormalised Polyakov loop. 
} 
\label{tab-1}     
\end{table}

We have available several ensembles, four below and four above the deconfinement transition, see Table~\ref{tab-1}. 
These ensembles make up our {\em Generation 2} ensembles and have been used in studies of  transport \cite{Amato:2013naa, Aarts:2014nba}, bottomonium  \cite{Aarts:2014cda}, open and hidden charm  \cite{Kelly:2018hsi}, and baryons \cite{Aarts:2015mma,Aarts:2017rrl}, which are further studied in this paper.
Tuning of the lattice parameters and also the ``zero-temperature'' ensemble ($N_\tau=128$) were kindly provided by the HadSpec collaboration \cite{Edwards:2008ja}.
The crossover temperature, denoted with $T_c$, is determined via the inflection point of the renormalised Polyakov loop and is found to be $T_c=185(4)$ MeV. This is higher than in nature, due to the light quarks  not having their physical masses ($m_\pi=384(4)$ MeV). The strange quark mass has its physical value \cite{Aarts:2014cda,Aarts:2014nba}.
We note here that in the case of a crossover the transition temperature determined via other observables, e.g.\ linked to chiral properties, can be different, see e.g.\ Ref.\ \cite{Borsanyi:2010bp}. Here we use $T_c$ to denote the transition temperature determined via the Polyakov loop, as discussed above, and take care to distinguish it from estimates coming from other observables.

For the baryonic correlators, we apply Gaussian smearing \cite{Gusken:1989ad} at both the source and the sink in the spatial directions, in order to increase the overlap with the groundstate.  The smearing parameters are chosen to maximise the length of the plateau for the effective mass of the groundstate at the lowest temperature.  These smearing parameters are then used at all temperatures, see Refs.\ \cite{Aarts:2015mma,Aarts:2017rrl}.


\section{In-medium effects in the hadronic phase}
\label{sec:had}

We computed all octet (spin 1/2) and decuplet (spin 3/2) baryon correlators, for both positive and negative parity. In fact, in each channel the parity partners are encoded in the same euclidean correlator: if $G_\pm(x)$ denotes the correlator projected to positive (negative) parity, i.e.,
\be
 G_\pm(x) = \left\bra \tr P_\pm O(x)\overline O(0) \right\ket, \qquad P_\pm = \half \left(\id \pm \gamma_4\right),
\ee
with $O(x)$ the baryon annihilation operator, then
\be
\label{eq:Gpm}
G_\pm(1/T-\tau,\pv) = -G_\mp(\tau,\pv),
\ee
i.e.\ the parity partner propagates from the opposite side of the euclidean lattice; see Ref.\ \cite{Aarts:2017rrl} for a detailed derivation.

 In the confined phase, we find that the correlators can be described by combinations of exponentials, allowing us to determine the groundstate masses $m_\pm$ as a function of the temperature. Examples of correlators and a description of fitting methods can be found in Ref.\ \cite{Aarts:2017rrl}; we follow the same approach here.
 Our results for the ground state masses, $m_\pm(T)$, in both parity channels at four temperatures in the confined phase are given in Table  \ref{tab:mass}, together with the ``zero-temperature'' results from the Particle Data Group \cite{Tanabashi:2018oca}. As mentioned above, the results for the $N$, $\Delta$ and $\Omega$ baryons were previously presented in Ref.\ \cite{Aarts:2017rrl}.

\begin{widetext}

\begin{table}[t]
\centering
\begin{tabular}{c  cc cccc || c}
\hline
$\;\;S\;\;$ & & $\;\; I(J^P) \;\;$ & $\;T/T_c= 0.24\;$ & 0.76 & 0.84 & 0.95 & PDG \\
\hline
\multirow{4}{*}{$0$} 	& $\;\;$\multirow{2}{*}{$N$} $\;\;$	& $\half(\half^+)$  & 1159(13) & $\;\;$ 1192(39) $\;\;$ & $\;\;$1169(53) $\;\;$  & $\;\;$1104(40) $\;\;$ & 939 \\
				& 					&  $\half(\half^-)$ & 1778(52) & 1628(104) & 1425(94) & 1348(83) & 1535 \\
				&  \multirow{2}{*}{$\Delta$}  & $\thalf(\thalf^+)$ & 1459(58) & 1521(43) & 1449(42) & 1377(37) & 1232 \\ 
				& 					 & $\thalf(\thalf^-)$ & 2138(117) & 1898(106) & 1734(97) & 1526(74) & 1710\\
\hline
				&  \multirow{2}{*}{$\Sigma$} & $1(\half^+)$ & 1277(13) & 1330(38) & 1290(44) & 1230(33) & 1193\\
				& 					 & $1(\half^-)$ &1823(35)  & 1772(91)  & 1552(65) & 1431(51)  & 1750 \\
\multirow{2}{*}{$-1$}	&  \multirow{2}{*}{$\Lambda$}  & $0(\half^+)$ &1248(12) & 1293(39)  &  1256(54) & 1208(26)  & 1116 \\
				& 					& $0(\half^-)$ & 1899(66)  &1676(136)  & 1411(90)  & 1286(75) & 1405--1670\\
				&  \multirow{2}{*}{$\Sigma^*$}  & $1(\thalf^+)$ & 1526(32) & 1588(40)  & 1536(43) & 1455(35) & 1385 \\
				& 					& $1(\thalf^-)$ & 2131(62) & 1974(122) & 1772(103) & 1542(60)  & 1670--1940\\
\hline
\multirow{4}{*}{$-2$}	& \multirow{2}{*}{$\Xi$}  	& $\half(\half^+)$ & 1355(9) &  1401(36) & 1359(41) & 1310(32) & 1318 \\
				& 					& $\half(\half^-)$ & 1917(27)  &  1808(92) & 1558(76) & 1415(50) & 1690--1950\\
				& \multirow{2}{*}{$\Xi^*$}  & $\half(\thalf^+)$  & 1594(24) & 1656(35) & 1606(40) & 1526(29) & 1530\\
				& 					& $\half(\thalf^-)$  & 2164(42)  & 2034(95) & 1810(77) & 1578(48) & 1820\\
\hline
\multirow{2}{*}{$-3$}	& \multirow{2}{*}{$\Omega$}   &  $0(\thalf^+)$ & 1661(21) & 1723(32) & 1685(37) & 1606(43) & 1672 \\
				& 					 & $0(\thalf^-)$ & 2193(30) & 2092(91) & 1863(76) & 1576(66) & 2250 \\
\hline
\end{tabular}
\caption{Groundstate masses $m_\pm$ (in MeV) for baryons with strangeness $S$ in both parity sectors ($P=\pm$) in the confined phase. Estimates for statistical and systematic uncertainties are included. The final column shows the $T=0$ values from the PDG. Note that in some cases there is more than one candidate.
}
\label{tab:mass}
\end{table}

\end{widetext}

A few things can be noted. We start at the lowest temperature. Since the light quarks are somewhat heavy, the $S=0$ states at the lowest temperature are also heavier than in nature. However, since  in our simulations the $s$ quark has its physical mass, for hyperons this difference is reduced as strangeness decreases. Negative-parity states are typically about 500-600 MeV heavier than their partners, both in our simulations and in the PDG. Some negative-parity states in the PDG seem anomalously light, such as the $\Lambda(1405)$, and the status of this state is indeed under discussion (see e.g.\ 
the review \cite{Hyodo:2011ur} and references therein).  In  these cases, Table \ref{tab:mass} also lists masses from the PDG which are separated by about 500 MeV and hence are potential candidates for parity partners, as suggested by our results at the lowest temperature (we note here that our spectroscopy methods are not specifically designed for high-precision spectroscopy in vacuum). As a final remark at the lowest temperature, we note that the positive-parity masses satisfy, to high precision, the Gell-Mann--Okubo mass relation \cite{GellMann:1961ky,Okubo:1961jc}
\be 
\label{eq:GMO1}
 \frac{3}{4} m_\Lambda + \frac{1}{4} m_\Sigma - \frac{1}{2}\left( m_N + m_\Xi\right) = 0,
 \ee
 for octet baryons and Gell-Mann's equal spacing rule
 \be
\label{eq:GMO2}
m_{\Sigma^*} - m_\Delta  = m_{\Xi^*} - m_{\Sigma^*} = m_\Omega - m_{\Xi^*}
\ee
 for decuplet baryons, also for our choice of quark masses, but the negative-parity masses do not (as is expected).

\begin{figure}[t]
\centering
\includegraphics[width=0.48\textwidth]{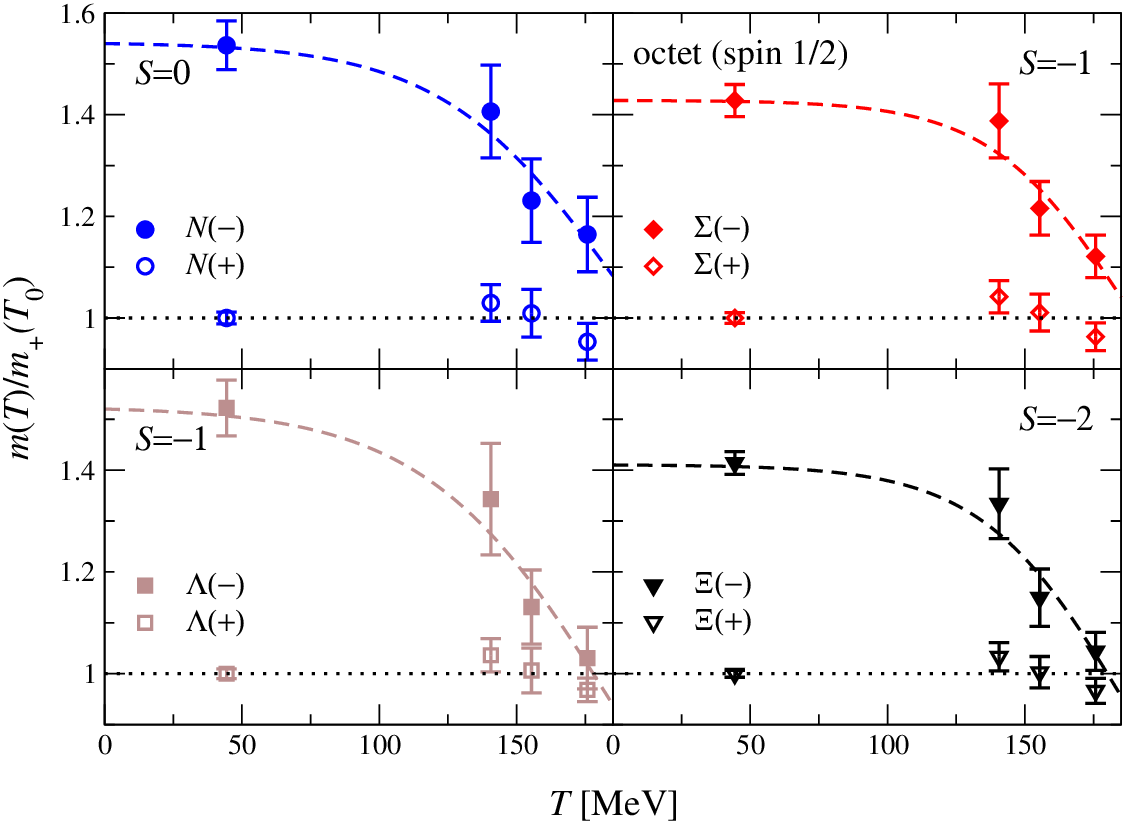}
\caption{Temperature dependence of the groundstate masses, normalised with $m_+$ at the lowest temperature, $m_\pm(T)/m_+(T_0)$, in the hadronic phase, for octet baryons. Positive- (negative-) parity masses are indicated with open (closed) symbols.
}
\label{fig:mass-octet}      
\end{figure}

\begin{figure}[t]
\centering
\includegraphics[width=0.48\textwidth]{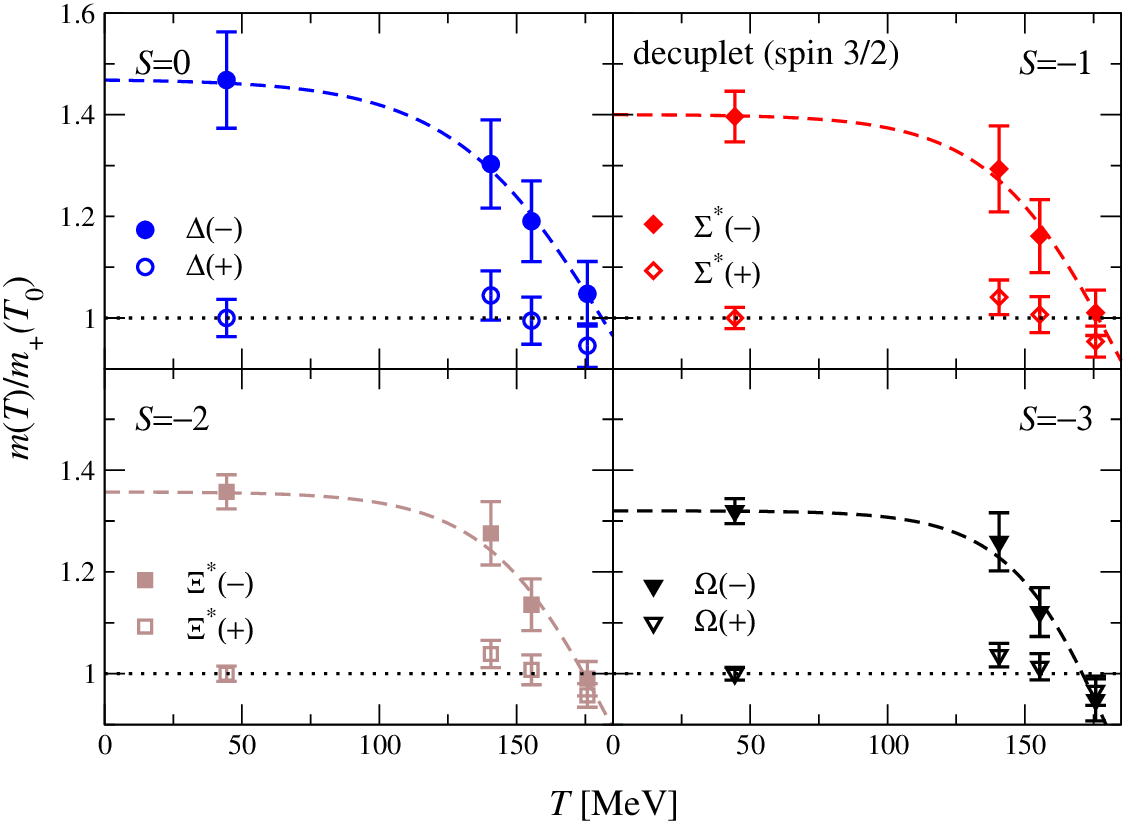}
\caption{As above, for decuplet baryons.}
\label{fig:mass-decuplet}      
\end{figure}

We now turn to the discussion of temperature effects, also presented in Table~\ref{tab:mass} and summarised in Figs.\ \ref{fig:mass-octet} and \ref{fig:mass-decuplet}, where we show $m_\pm(T)$ in the various channels, normalised with $m_+$ at the lowest temperature, $T_0=44$ MeV. 
Several  observations can be made. The positive-parity masses are largely temperature independent. A slight increase and subsequent drop when approaching the transition can be seen, but it is not significant within current errors. A corollary is that the relations (\ref{eq:GMO1}, \ref{eq:GMO2}) are satisfied throughout the confined phase (within error), which constrains thermal model-building efforts. 
The negative-parity masses on the other hand drop in all channels in a similar way, and become near-degenerate with the corresponding positive-parity mass near the transition. The larger errors for the negative-parity states indicate that it is harder to fit the negative-parity correlators, impeding very precise statements about the degeneracy very close to the transition.

To model the temperature dependence of $m_-(T)$, we have fitted the data with a simple Ansatz, interpolating between $m_-(0)$ and $m_-(T_c)$,
\be
\label{eq:1}
m_-(T) = w(T,\gamma)m_-(0) + \left[1 - w(T,\gamma)\right]m_-(T_c).
\ee
Here  
\be
w(T,\gamma) = \frac{\tanh\left[\gamma^{-1} (1 - T/T_c)\right]}{\tanh( \gamma^{-1} ) }
\ee
is a transition function, with $w(0,\gamma)=1$ and $w(T_c,\gamma) = 0$. 
Since the deconfinement and chiral transitions need not coincide, a more elaborate Ansatz would replace $T_c$ with $T_c^\chi$, a chiral transition temperature determined by e.g.\ the requirement that $m_-(T_c^\chi)=m_+(T_c^\chi)$, i.e.\ the temperature at which the masses become degenerate. Since there are only 4 data points in the confined phase and the uncertainty is largest close to $T_c$, we choose to proceed here with the Ansatz (\ref{eq:1}), postponing to Sec.\ \ref{sec:decon} an estimate of $T_c^\chi$ directly from the euclidean correlators in each channel.

Since the transition is a crossover rather than a real phase transition, with the strength of the transition depending on the masses of the light quarks, the fit parameter $\gamma>0$ is used to encode the width of the transition region,  with small (large) $\gamma$ corresponding to a narrow (broad) crossover region. We have carried out fits in each of the 8 channels and find 
\be
0.22\lesssim \gamma\lesssim 0.35, \qquad 0.85\lesssim \frac{m_-(T_c)}{m_+(0)} \lesssim 1.1.
\ee
 The largest uncertainty resides in $m_-(T_c)$, since it assumes that the concept of a well-defined groundstate at or close to $T_c$ remains sensible. Note that in the fits we fixed $m_-(0)=m_-(T_0)$. The results of the fits are shown as dashed lines in  Figs.\ \ref{fig:mass-octet} and \ref{fig:mass-decuplet}. We use the results of this analysis in the next section.

To summarise, our results confirm that parity doubling indeed emerges, as expected, and moreover establish that the degeneracy develops from a reduction of the negative-parity masses with temperature, while the positive-parity masses remain approximately constant. We find this to be similar in all the channels studied. While the emergence of degeneracy is not surprising, the manner in which this is realised cannot be determined from symmetry alone, but requires a dynamical calculation, as we have done here. 
We note that our findings can be used to constrain effective parity-doublet models \cite{Detar:1988kn,Jido:1999hd,Jido:2001nt}, in which a chirally invariant component to baryon masses, denoted as $m_0$, is introduced. These models permit a nonzero baryon mass also when chiral symmetry is restored, with the typical behaviour of the mass of parity partners being  \cite{Detar:1988kn,Jido:1999hd,Jido:2001nt,Steinheimer:2011ea}
\be
m_\pm= \sqrt{m_0^2+ c_1\sigma_0^2} \mp c_2 \sigma_0,
\ee
where $\sigma_0$ denotes the chiral condensate in vacuum and $c_{1,2}$ are parameters related to the couplings between the baryons and the light mesons. Incorporating medium effects by allowing $\sigma_0$ to be temperature dependent yields a prediction for the temperature dependence of the baryon masses, but some nontrivial interplay is to be expected to ensure that $m_+$ remains largely independent of temperature, as our data indicates.
Some recent work along these lines can be found in Refs.\ \cite{Benic:2015pia,Motohiro:2015taa,Nishihara:2015fka,Suenaga:2017wbb,Takeda:2017mrm,Mukherjee:2017jzi,Sasaki:2017glk,Yamazaki:2018stk} (see also Ref.\ \cite{Chen:2017pse} for an approach based on the Faddeev kernel).
Studies determining only the positive-parity hyperon groundstates at nonzero temperature include Ref.\ \cite{Torres-Rincon:2015rma}, using the Polyakov-Nambu-Jona-Lasinio (PNJL) model, and Ref.\ \cite{Azizi:2016ddw}, using QCD sum rules.


\section{In-medium hadron resonance gas}
\label{sec:HRG}

  We now consider a possible application of our findings in the confined phase, which has relevance for heavy-ion phenomenology. 
 A widely used model to describe thermodynamical properties of QCD at low temperature is the hadron resonance gas (HRG), in which all resonances identified in the PDG contribute to the pressure and other thermodynamic quantities, such as generalised susceptibilities. In the standard formulation, the resonances are not interacting and their properties are taken as in vacuum, i.e.\ not affected by the presence of a (thermal) medium. This ideal HRG gives a reasonable description of experimental and lattice QCD data, but discrepancies appear at a quantitative level. As mentioned in the Introduction, several modifications of the HRG have been proposed to cure this \cite{Vovchenko:2016rkn,Huovinen:2017ogf,Satarov:2016peb}.
Since the ideal HRG results typically fall below e.g.\ the lattice QCD data for thermodynamic quantities, it has been proposed to include more states than have been identified in the PDG, since this will result in an increase of the pressure and of susceptibilities. Indeed, including states which appear in the quark model but have not (yet) been identified experimentally leads to a better agreement between  lattice data and this extended HRG \cite{Majumder:2010ik,Bazavov:2014xya,Alba:2017mqu}.

Motivated by our findings above, we propose here to include medium effects in the hadronic masses, rather than keeping them as in vacuum. Since we observed that the negative-parity groundstate masses drop as the temperature increases, we can anticipate an increase in the pressure and susceptibilities compared to the ideal HRG, since lighter degrees of freedom lead to larger fluctuations. Hence a qualitative improvement is expected. We will refer to this proposal as the in-medium HRG, to distinguish it from other modifications.
Since in this paper we are concerned with spectral changes for baryons, we apply the in-medium HRG to quantities that are sensitive to baryon number, namely the baryonic contributions to the pressure and the correlation between baryon number and strangeness, 
\be
\chi_{BS} = \frac{1}{VT}\bra BS\ket = \frac{T}{V} \frac{\partial^2\ln Z}{\partial\mu_B\partial\mu_S},
\ee
where $V$ is the spatial volume and $Z$ the partition function.

\begin{figure}[t]
\centering
\includegraphics[width=0.45\textwidth]{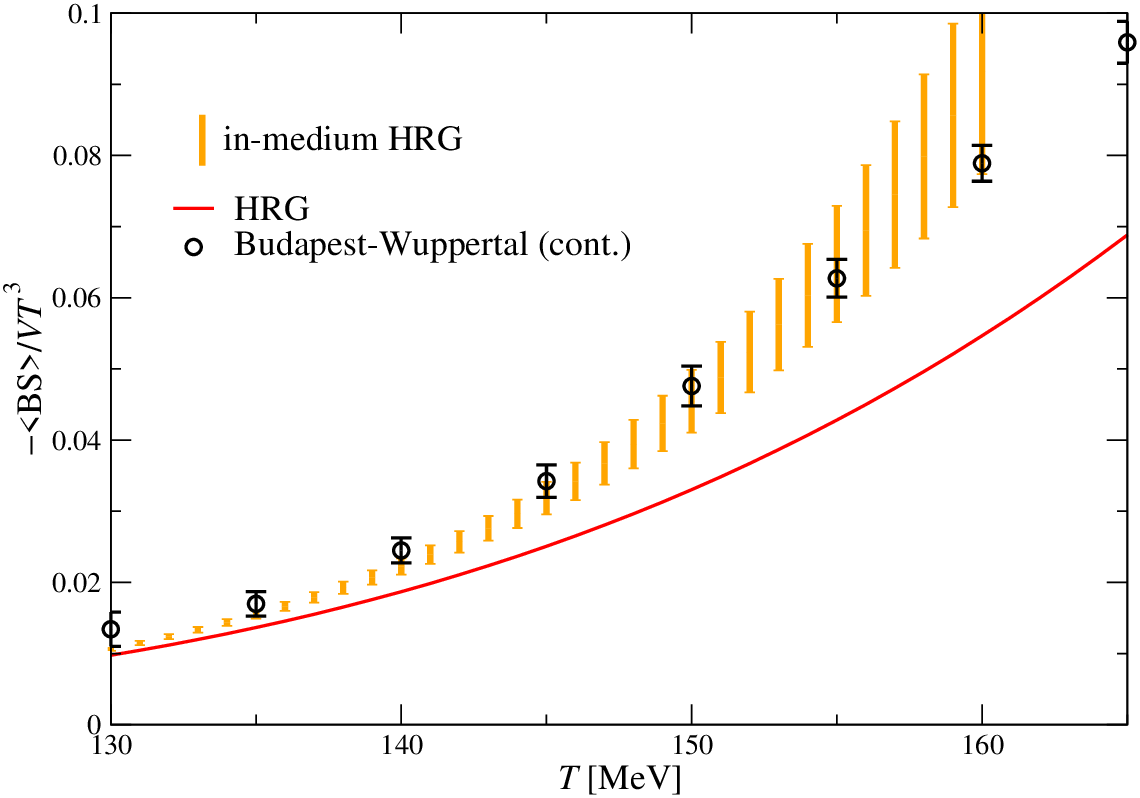}
\caption{Fluctuations of strange baryons,  $-\chi_{BS}/T^2 = -\bra BS \ket/VT^3$.
The symbols are continuum-extrapolated lattice data from the Budapest-Wuppertal collaboration \cite{Borsanyi:2011sw,Bellwied:2015lba,Borsanyi:priv}.
HRG (full line) is the standard hadron resonance gas. The in-medium HRG (orange stripes) uses temperature-dependent masses for the negative-parity groundstates, and is explained in the text.
}
\label{fig:3}       
\end{figure}

\begin{figure}[t]
\centering
\includegraphics[width=0.48\textwidth]{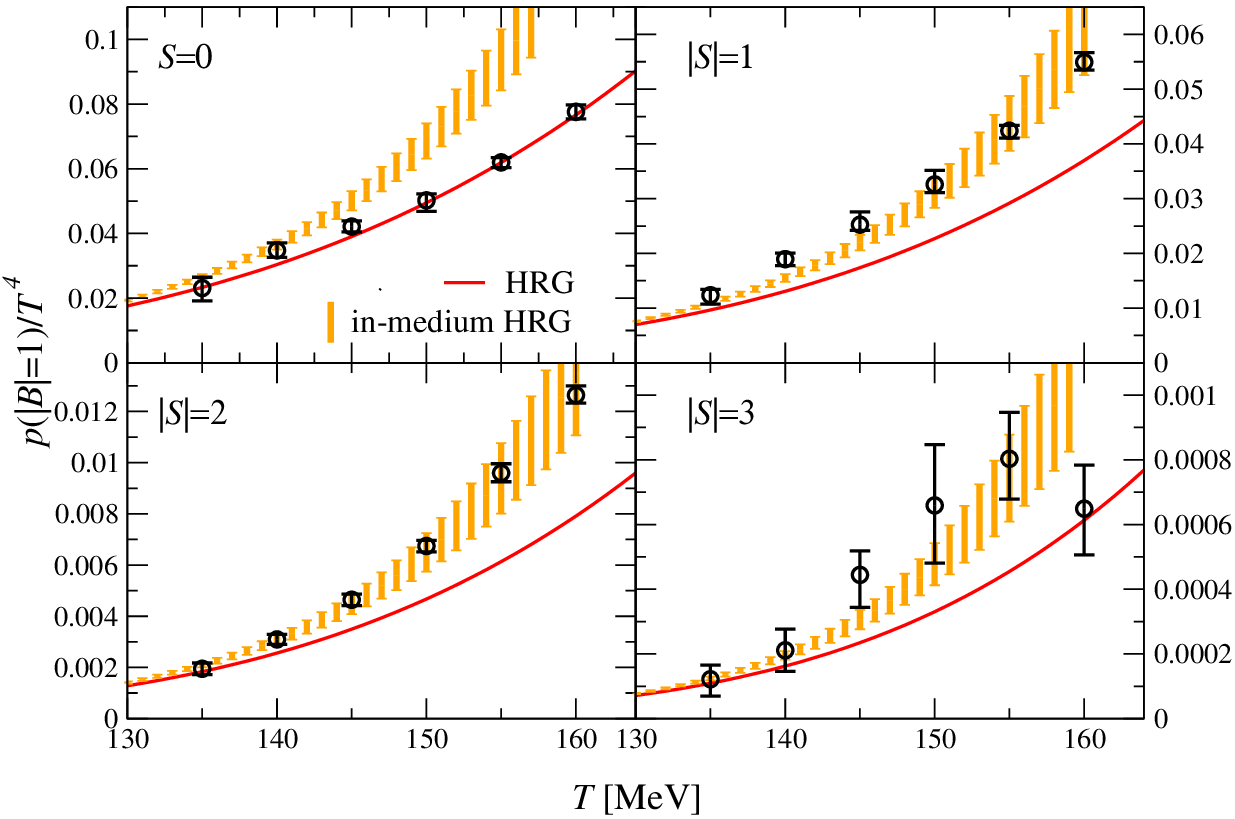}
\caption{Contributions to the normalised pressure $p/T^4$ from the sectors with baryon number $|B|=1$ and strangeness $|S|=0,1,2,3$. The lattice data is taken from Ref.\ \cite{Alba:2017mqu,Ratti:priv}. The HRG and in-medium HRG are as above.
}
\label{fig:4}      
\end{figure}

Let us now turn to the concrete implementation. As supported by the lattice study, positive-parity masses are kept fixed, i.e.\ at their vacuum value, but the negative-parity groundstate masses are taken to be temperature dependent, according to the prescription (\ref{eq:1}). We do not modify masses of excited states, as we have currently no predictions for their temperature dependence from the lattice, but one could envision including this. Our lattice study was carried out using two light flavours which are heavier than in nature. In our in-medium HRG application, we use the actual mass values given in the PDG: concretely, we use the PDG2016 baryon masses classified with 3 and 4 stars, up to 2.5 GeV. In the case that there are several candidates for negative-parity partners -- e.g.\ $\Lambda(1450)$ and  $\Lambda(1670)$ -- we have identified as parity partners the ones that are separated by about 500-600 MeV,  i.e.\ $\Lambda(1670), \Sigma(1940)$ and $\Xi(1950)$, similar as for the other baryons. For the pseudocriticial temperature, we used $T_c=155$ MeV,  which is determined by chiral observables \cite{Borsanyi:2010bp,Bazavov:2011nk}. 
The strength of the transition is expected to depend  on the masses of the light flavours. To incorporate this, we have varied the parameter $\gamma$ encoding the width of the transition region, see Eq.\ (\ref{eq:1}), but have found no effect within the uncertainty arising from varying $m_-(T_c)/m_+(0)$.
For the results shown in Figs.\ \ref{fig:3} and \ref{fig:4}, we have used $\gamma = 0.3$ and varied the ratio $m_-(T_c)/m_+(0) $ between 1 and 1.1.

In Fig.\ \ref{fig:3},  we show the dimensionless combination $-\chi_{BS}/T^2 = -\bra BS \ket/VT^3$, representing fluctuations of strange baryons. The standard HRG result, shown by the full line, lies below the continuum-extrapolated lattice data from the Budapest-Wuppertal collaboration \cite{Borsanyi:2011sw,Bellwied:2015lba,Borsanyi:priv}. The in-medium HRG result moves away from the HRG line as the temperature increases, since a subset of the strange baryons becomes lighter as the system heats up, in qualitative agreement with the lattice data. Within the uncertainty, which is dominated by the variation of $m_-(T_c)/m_+(0)$, we find that the in-medium HRG result agrees with the lattice data also quantitatively.

Fig.\ \ref{fig:4} shows the contributions to the normalised pressure $p/T^4$ from the sectors with baryon number $|B|=1$, organised by strangeness. The ideal and in-medium HRG are compared to the lattice data of Refs.\ \cite{Alba:2017mqu,Ratti:priv}. In the $S=0$ sector, the ideal HRG describes the data very well and in fact modifications of the HRG will typically make the comparison worse.
In the $|S|=1,2,3$ sectors, however, the standard HRG lies significantly below the lattice data.  Here we observe that the in-medium HRG leads to a quantitative improvement.

To conclude, a few comments are in order. First we note that both mesons and baryons enter in the HRG and for many thermodynamic quantities mesons, containing the lightest states, dominate. In our lattice study, we have considered baryons only. Hence in the application above we have consider quantities that are sensitive to baryon number. To extend the in-medium HRG to include mesonic resonances requires a detailed study the mesonic spectrum as the temperature increases, which we leave for the future. 
Second, with temperature-dependent masses, an application of the in-medium HRG to thermodynamic quantities which are obtained as temperature derivatives of the pressure, such as the entropy density $s=\partial p/\partial T$,  care has to be taken to be thermodynamically consistent, see e.g.\ Ref. \cite{Levai:1997yx}.
Finally, given that there are a number of HRG modifications which aim to improve the ideal HRG in complementary ways, it would be interesting to consider various modifications simultaneously, provided they are not in incompatible on physical grounds.


\section{Parity doubling}
\label{sec:decon}

We now return to the lattice data and analyse the transition from the hadronic phase to the quark-gluon plasma, in particular the emergence of parity doubling.

At $T=201$ MeV ($T/T_c=1.09$),  the lowest temperature in the deconfined phase we have access to, we find that clearly identifiable bound states do not appear to be present; evidence in the $N$ and $\Omega$ channels has already been reported in Ref.\  \cite{Aarts:2017rrl}. Instead, we study the emergence of parity doubling, which can be done directly at the level of the correlators $G_\pm$, without the need for identifiable groundstates. We consider the ratio \cite{Datta:2012fz,Aarts:2015mma}
\be
\label{eq:Rtau}
R(\tau) = \frac{G_+(\tau) - G_+(1/T-\tau)}{G_+(\tau) + G_+(1/T-\tau)},
\ee
to analyse the difference between the channels with opposite parity, see Eq.\ (\ref{eq:Gpm}).
Note that we consider zero momentum only and hence drop the $\vecp$ dependence.
In the case of parity doubling, $G_+(\tau) =G_+(1/T-\tau)=-G_-(\tau)$ and $R(\tau)=0$. In absence of parity doubling, and in the presence of clearly separated groundstates with a gap between the positive- and negative-partity groundstates, such that
\be
\label{eq:GG}
G_\pm(\tau) \sim \pm A_\pm e^{-m_\pm \tau}, \qquad m_- \gg m_+,
\ee
one finds that $R(\tau)=1$, in the interval where the groundstates dominate \cite{Datta:2012fz,Aarts:2015mma,Aarts:2017rrl}.
We can capture this in a single quantity $R$ by summing over the temporal lattice points,
\be
\label{eq:R}
 R = \frac{\sum_n R(\tau_n)/\sigma^2(\tau_n)}{\sum_n 1/\sigma^2(\tau_n)},
\ee
where $\sigma(\tau_n)$ denotes the error at timeslice $\tau_n$ and $4\leq n\leq N_\tau/2-1$, to exclude lattice artefacts and excited states present at early times. (Note that the same procedure was followed in Refs.\ \cite{Aarts:2015mma,Aarts:2017rrl}; the starting value in the sums in Refs.\ \cite{Aarts:2015mma,Aarts:2017rrl} was incorrectly given as $n=1$.)
 If the Ansatz of a single exponential at low temperature, Eq.\ (\ref{eq:GG}), describes the data well, one finds that the quasi-order parameter $R$ equals 1, while it vanishes in the case of parity doubling.

 \begin{figure}[t]
\centering
\includegraphics[width=0.48\textwidth]{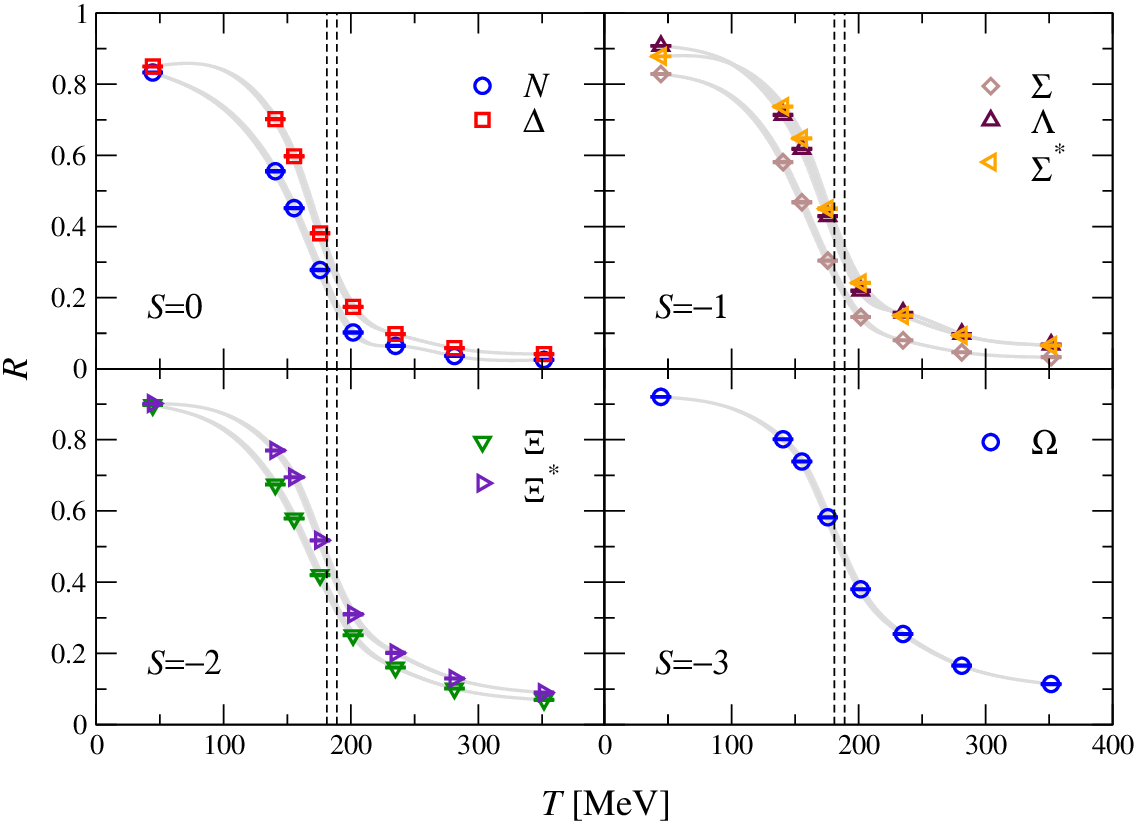}
\caption{Crossover behaviour of $R$, see Eqs.\ (\ref{eq:Rtau}, \ref{eq:R}), as a function of temperature and ordered by strangeness, indicating the emergence of parity doubling in the deconfined phase. The grey lines are cubic spline fits, with the width indicating a statistical uncertainty only. 
The vertical lines indicate the transition temperature extracted from the renormalised Polyakov loop, $T_c=185(4)$ MeV.
}
\label{fig:R}      
\end{figure}

 The results for $R$ as a function of temperature are shown in Fig.\ \ref{fig:R}. Note that the channels are identified by the particle content and are ordered by strangeness.
We note that $R$ indeed changes from (close to) 1 to (close to) 0 as $T$ increases, with the crossover taking place in the transition region as determined by the Polyakov loop.
 At the highest temperatures, the difference from 0 appears directly proportionate to the number of strange quarks in the channel, which reflects the explicit chiral symmetry breaking arising from the heavier strange quark. Eventually, this effect should disappear as $m_s/T\to 0$.
 We hence conclude that the interpretation of parity doubling due to chiral symmetry restoration in the baryon sector is indeed valid.

 \begin{figure}[t]
\centering
\includegraphics[width=0.48\textwidth]{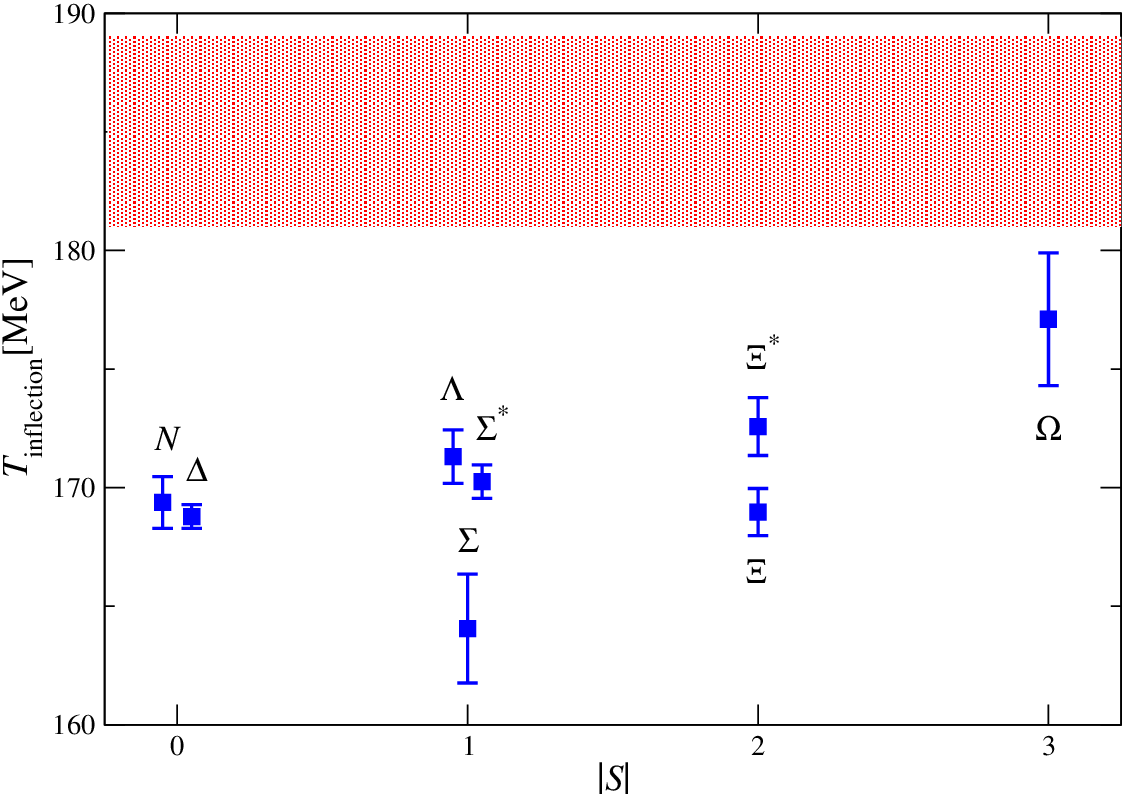}
\caption{
Temperature of the inflection points, $T_{\rm inflection}$[MeV], of the $R$ ratios shown in Fig.\ \ref{fig:R}, organised by strangeness (data points have been shifted horizontally for clarity). The horizontal band indicates the transition temperature extracted from the Polyakov loop.
}
\label{fig:R2}      
\end{figure}

When the transition between two phases is not a phase transition but merely an analytic crossover, as is the case for the thermal transition in QCD \cite{Aoki:2006we}, a single transition temperature cannot be given. Instead one finds a range of transition or pseudocritical temperatures, which depend on the observable and method used to define it, see e.g.\ Ref.\ \cite{Borsanyi:2010bp} and references therein. 
In Sec.\ \ref{sec:had} we denoted the transition temperature linked to parity doubling as $T_c^\chi$, but did not determine it. Here we will estimate it  directly from the $R$ ratio.
An often used method is to locate the inflection point of (quasi-)order parameters. For this reason, we have fitted cubic splines to the lattice data in Fig.\ \ref{fig:R}, which are represented by the grey lines, with the widths giving an indication of the statistical error. Using the cubic splines, we have extracted the temperatures of the inflection points, which are shown in Fig.\ \ref{fig:R2}, organised by strangeness and labeled by the groundstate of the respective channels. The errors are statistical only and determined by the bootstrap method.
The transition temperature determined from the renormalised Polyakov loop, $T_c=185(4)$ MeV, is indicated by the shaded band. 
We observe that the inflection-point temperatures lie below this band, by about  9 to 15 MeV (the result in the $\Sigma$ channel lies somewhat lower than expected, see also Fig.\  \ref{fig:R}). Due to the crossover nature of the transition, this is not unexpected.
A possible dependence on strangeness (see e.g.\ Ref.\ \cite{Bellwied:2013cta} on flavour separation during the crossover) might become better visible when the light quarks are tuned closer to their physical values. This project is currently in progress \cite{Jonas:2018}.


\section{Summary}
\label{sec:conc}

In this paper we determined the response of hyperons to an increase of temperature in thermal QCD, going from the hadronic phase to the quark-gluon plasma, using lattice QCD simulations.  While the masses of the positive-parity groundstates were seen to be not affected, within the numerical uncertainty, the masses of the negative-parity groundstates showed a characteristic temperature dependence, in all channels. We have argued that both findings are relevant for the phenomenology of heavy-ion collisions, in particular in the context of the hadron resonance gas. We have formulated an in-medium HRG to incorporate the temperature dependence of the negative-parity groundstates and applied it to fluctuations involving strange baryons and baryonic contributions to the pressure, which had been computed independently on the lattice. While in the nucleon and $\Delta$ ($S=0$) channels the standard HRG  describes the lattice data already very well (and any modification makes the comparison worse), we found that for hyperons the in-medium HRG leads to an improved agreement between lattice QCD results for those quantities and the HRG. It would therefore be interesting to extend this approach to mesons as well and determine the temperature dependence of spectral features of mesons. This work is currently in progress. Besides this, it would be of interest to combine our suggestion for the in-medium HRG with other modifications of the ideal HRG, to be able to study how various modifications may work together. Finally, our results in the hadronic phase can also be used to benchmark effective models, such as parity doublet and PNJL-type models.

At the higher temperatures, proceeding into the quark-gluon plasma, we focussed on the emergence of parity doubling, at the level of the correlators. We defined an $R$ ratio, which captures the transition from the chirally broken to the chirally symmetric phase. The temperatures of the inflection points of this $R$ ratio lie somewhat below the pseudocritical temperature extracted from the Polyakov loop, indicating the crossover nature of the transition. To observe possible strangeness dependence, a larger separation between the strange and light quark masses would be advantageous. 
This would also allow us to further study modifications to the hadronic spectrum in the low-temperature phase, closer to the physical point.
Work in this direction is currently underway.


\vspace*{0.5cm}

\acknowledgments

We thank Szabolcs Bors\'anyi  and Claudia Ratti for providing the lattice data shown in Figs.\ \ref{fig:3}  and \ref{fig:4} respectively, and Rene Bellwied for suggesting the analysis of the inflection points. 
We thank Simon Hands and Jonivar Skullerud for collaboration and encouragement.
We are grateful for support from STFC via grants ST/L000369/1 and ST/P00055X/1, the Swansea Academy for Advanced Computing (SA2C), and ICHEC.
This work has been performed in the framework of COST Action CA15213 THOR.
We acknowledge PRACE for access to the Marconi-KNL system hosted by CINECA, Italy. 
Computing resources were made available by HPC Wales and Supercomputing Wales, and by the DiRAC Blue Gene Q Shared Petaflop system at the University of Edinburgh, operated by the Edinburgh Parallel Computing Centre on behalf of the STFC DiRAC HPC Facility (www.dirac.ac.uk). This equipment was funded by BIS National E-infrastructure capital grant ST/K000411/1, STFC capital grant ST/H008845/1, and STFC DiRAC Operations grants ST/K005804/1 and ST/K005790/1. DiRAC is part of the National E-Infrastructure.


\end{document}